\documentclass{kapproc} % Computer Modern font calls

\usepackage{t1enc}

\usepackage{procps} 

\usepackage[dvips]{graphicx}

\upperandlowercase

\kluwerbib % will produce this kind of bibliography entry:

\let\lcitebracket(
\let\rcitebracket)

\begin{document}

\articletitle[Star Formation in Three Nearby Galaxy Systems]
{Star Formation in Three Nearby Galaxy Systems}
\author{S. Temporin,\altaffilmark{1} S. Ciroi,\altaffilmark{2}
A. Iovino,\altaffilmark{3} E. Pompei,\altaffilmark{4}
M. Radovich,\altaffilmark{5},
and P. Rafanelli\altaffilmark{2}}

\affil{\altaffilmark{1}Institute of Astrophysics, University of Innsbruck, \ 
\altaffilmark{2}Astronomy Department, University of Padova, \ 
\altaffilmark{3}INAF-Brera Astronomical Observatory, \
\altaffilmark{4}ESO-La Silla, \
\altaffilmark{5}INAF-Capodimonte Astronomical Observatory}

\begin{abstract}
We present an analysis of the distribution and strength of
star formation in three nearby small galaxy systems, which are undergoing
a weak interaction, a strong interaction, and a merging process, respectively.
The galaxies in all systems present widespread star formation enhancements,
as well as, in some cases, nuclear activity.
In particular, for the two closest systems, we study the number-count,
size, and luminosity distribution of H\,{\sc ii} regions within the interacting
galaxies, while for the more distant, merging system we analyze the general
distribution of the H$\alpha$ emission across the system and its velocity field.
\end{abstract}


\section{Introduction}
Galaxy interactions have been known for a long time to trigger star formation,
although both observations and numerical simulations have shown that the
enhancement of star formation depends, among other factors, on the orbital
geometry of the encounter. In some situations interactions might even
suppress star formation. Hence, the star formation properties of interacting
systems may serve as a clue to their interaction history.

Here we analyse the star formation properties of three nearby galaxy systems
in differing evolutionary phases: the weakly interacting triplet AM 1238-362
\cite{T03a}, the strongly interacting compact galaxy group SCG 0018-4854 \cite{I02,OI97},
and the merging compact group CG J1720-67.8 \cite{W99}.
The first two systems are at comparable distances (48 and 45.5 Mpc, respectively, 
assuming H$_0$ = 75 km s$^{-1}$ Mpc$^{-1}$), close enough to allow us to study the
luminosity and size distribution of the H\,{\sc ii} regions in the individual 
galaxies.
The third system, at a distance of 180 Mpc, does not offer us the possibility
to study the properties of individual H\,{\sc ii} regions. However, the general
distribution of the ionized gas across the whole group and its velocity field
can be used to guess the evolutionary status of the group. 
We interpret our (in some cases still preliminary) results taking into account
star formation properties of other nearby galaxies \cite{Th02} and 
numerical simulations from the literature (e.g. Mihos, Bothun, \& Richstone 1993). 

\section{AM 1238-362}

\begin{figure}[hbt]
\centerline{
\hbox{
\includegraphics[height=6cm]{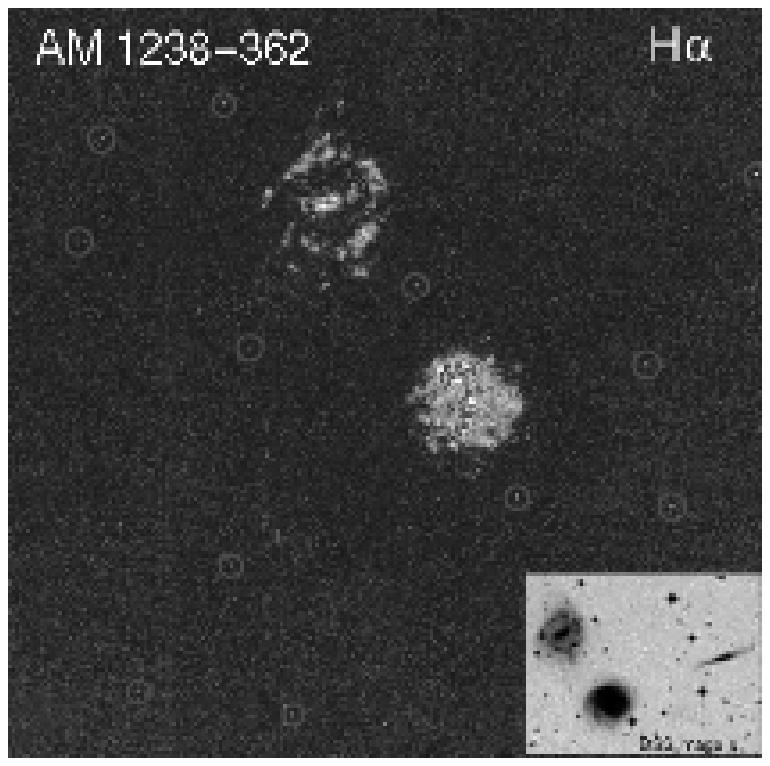}
\includegraphics[width=6cm]{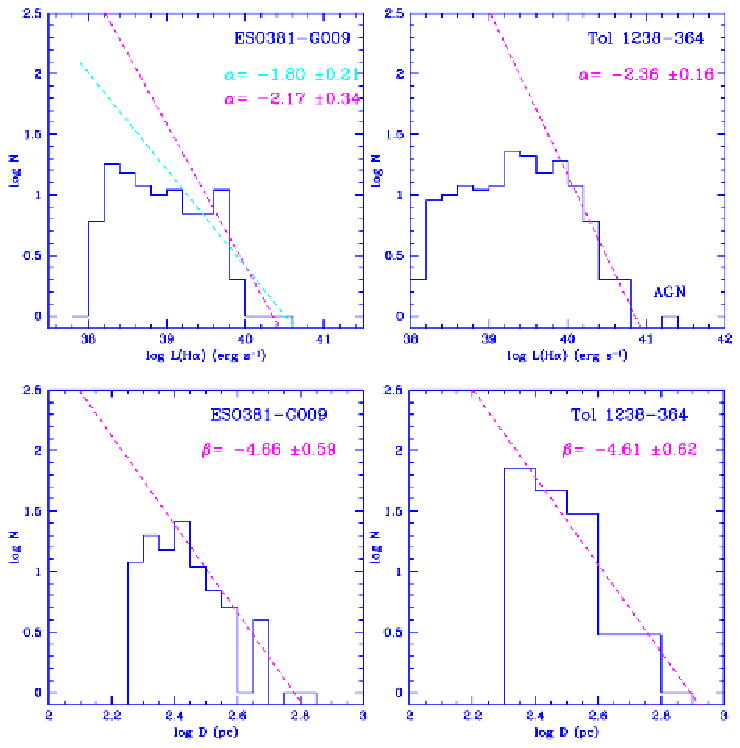}}}
\sidebyside
{\caption{Continuum-subtracted H$\alpha$ image of
ESO 381-G009 (NE) and Tol~ 1238-36.4 (SW). North is up, East to the left. 
Emission-line point-like sources are circled in red. The insert in the lower-right
corner shows a DSS image of the triplet. \label{AM_a}}}
{\caption{H\,{\sc ii} region luminosity functions and size distributions for
ESO 381-G009 and Tol1238-36.4. The magenta lines represent the fitting laws.
The cyan line represents an alternative fit obtained on a wider luminosity
range ($\log$(L$_{\rm{H\alpha}}$) $>$ 39).}\label{AM_b}}
\end{figure}

AM 1238-362 is a fairly tight triple system composed by two nearly face-on
(ESO 381-G009 and Tol 1238-36.4) and an edge-on (ESO 381-G006)
spiral galaxies (Fig.~\ref{AM_a}). A detailed study of the morphological and
spectrophotometric properties of the two face-on galaxies has
been presented in Temporin et al. (2003a), where the interaction signatures
and star formation properties have been discussed in the framework
of the evolutionary history of the system.  
The two galaxies exhibit enhanced star formation activity, and the radial and 
azimuthal distribution of their H$\alpha$ emission show a tendency for both 
disks to concentrate the star formation activity on the side facing the companion
(see Temporin et al. 2003a, Fig. 20). 
Here we make use of
H$\alpha$ narrow-band imaging obtained at the ESO-MPIA 2.2 m telescope
(see the above cited work for details on observations and data reduction) to
further investigate the star formation properties of these mildly interacting
galaxies. In particular we measured the H$\alpha$ luminosities and the sizes of
the individual H\,{\sc ii} regions in order to analyse their luminosity
functions (LFs) and size distributions. The regions were identified and measured
in automated way with the {\sc HIIphot} algorithm \cite{Th00},
choosing as a detection threshold a signal-to-noise ratio S/N $>$ 5$\sigma$.
The LFs are shown in Fig.~\ref{AM_b}. Weighted least square fits to the
regions in the luminosity range where LFs were believed to be complete were 
obtained assuming a standard power-law distribution with index $\alpha$. Similarly
we obtained power-law fits to the size distributions of the H\,{\sc ii} regions.
The slope $\beta$ of the size distribution is expected to be related
to that of the LF by the relation $\beta\,=\,2 + 3\alpha$ \cite{Oey03}.
For Tol 1238-36.4, a luminous infrared galaxy (LIRG) classified as an SB(rs)bc, 
we obtained $\alpha$ = $-$ 2.36$\pm$0.16, in good agreement with expectations 
for its morphological type \cite{K89}, and $\beta$ = $-$ 4.61$\pm$0.62,
in agreement (within errors) with the above relation. 
However, we observe a uniform shift of 
the LF towards higher luminosities by about one order of magnitude 
with respect to other nearby star-forming spiral galaxies (e.g. Thilker 
et al. 2002). This feature of the LF is typical of LIRGs
(see also Alonso-Herrero, these proceedings).
In this particular case, the high luminosity of the H\,{\sc ii} regions could
be a consequence of the increased star formation activity triggered by the 
interaction. Since the influence of the AGN is mostly confined within 
the central kpc \cite{T03a}, we rule out the possibility of a significant 
cotribution to the ionization by the central AGN. 
For ESO 381-G009, the slope of the LF strongly depends on the choice of the 
fitted luminosity range. The steeper slope ($\alpha$ = $-$2.17$\pm$0.34) 
shown in Fig.~\ref{AM_b}, obtained for regions with $\log$(L$_{\rm{H\alpha}}$) $>$ 39.5, 
is in good agreement with the slope of the size distribution.

The H$\alpha$ image revealed also the presence of 12 emission-line point-like sources
in the intergalactic space. Their projected distances to the nearest galaxy 
range between 12 and 68 kpc and their H$\alpha$ luminosities are in the range
0.6 - 7.4 $\times$ 10$^{38}$ ergs s$^{-1}$. These could be intergalactic H\,{\sc ii} regions
similar to those recently found in other interacting systems or around cluster galaxies
\cite{RW04}.

\section{SCG 0018-4854}
This strongly interacting galaxy group is composed by a tight quartet
\cite{I02} and a concordant redshift spiral galaxy (ESO 194-G13) $\sim$ 12 arcmin apart.
All members exhibit strong signs of disturbances, the most outstanding
being the $\sim$ 30 kpc-long tidal tail of NGC 92.
Here we present a preliminary analysis of the H$\alpha$ images of the 5 galaxies.
The observations were carried out in 2002 September and 2003 January at the
ESO-VLT-UT4 telescope equipped with FORS2. The H$\alpha$ images were reduced
in a standard way and flux-calibrated by means of a standard star observed 
during the same nights, taking into account the filter transmission curve and the
contribution from the [N\,{\sc ii}] $\lambda\lambda$ 6548, 6583 emission lines.
Appropriately scaled R-band images were used for the continuum subtraction.

\begin{figure}[hbt]
\centerline{
\vbox{
\hbox{
\includegraphics[height=7cm]{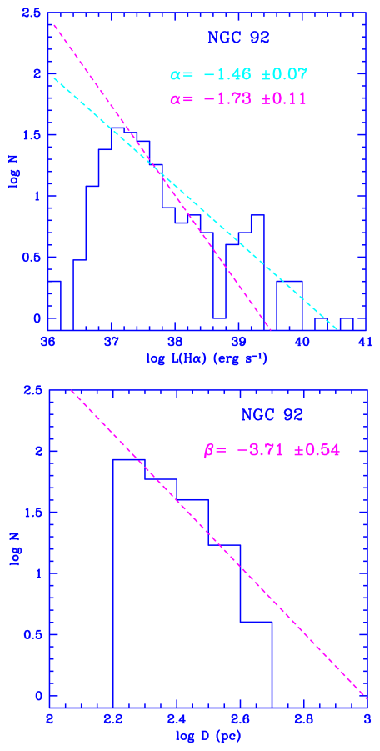}
\includegraphics[height=6.8cm]{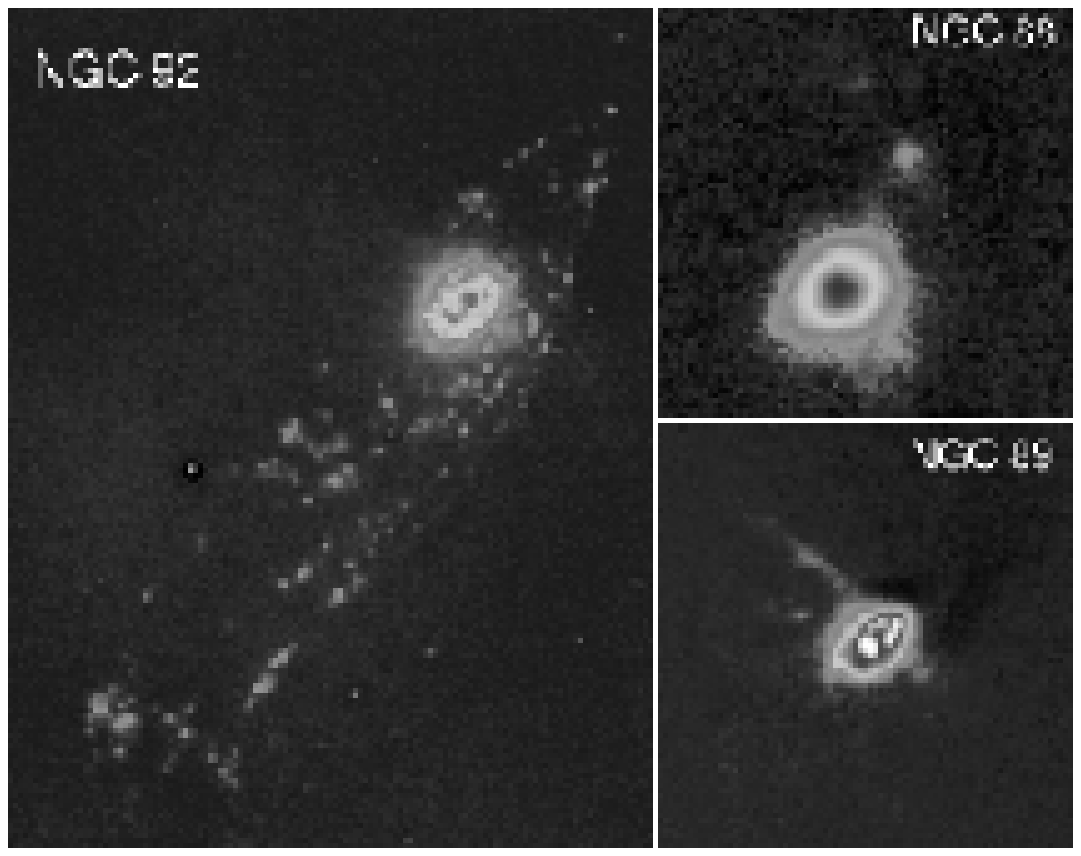}}
\includegraphics[width=12.4cm]{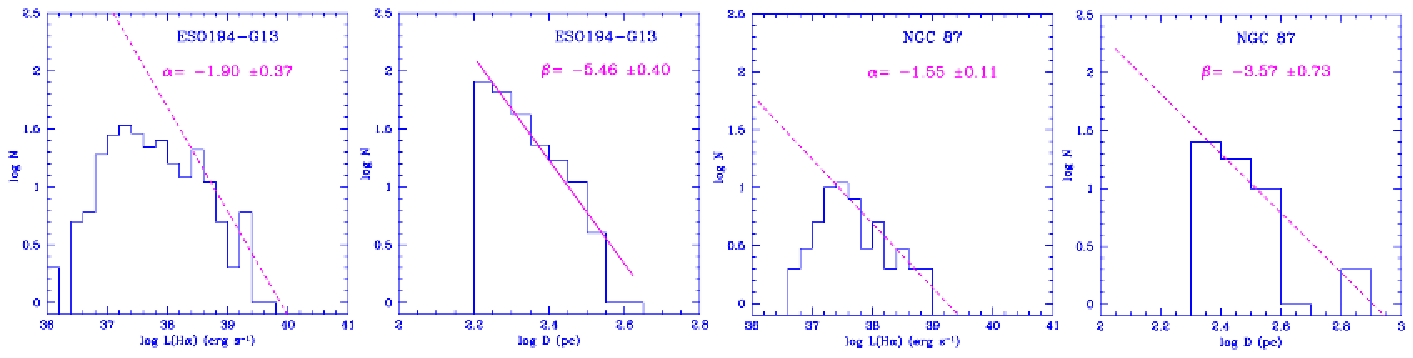}}}
\caption{\emph{Top left:} H\,{\sc ii} region LF and size distribution of NGC 92.
Two possible fits to the LF are shown, the flatter takes into account 
also the regions at the high luminosity end of the distribution. \emph{Top right:}
Continuum-subtracted H$\alpha$ VLT-FORS2 images of NGC 92, NGC 88, and
NGC 89. \emph{Bottom:} H\,{\sc ii} region LFs and size distributions of ESO 194-G13 
and NGC 87.} \label{LF_scg0018}
\end{figure}

In three of the galaxies, individual H\,{\sc ii} regions were detected in sufficient 
number to allow us the construction of LFs and size distributions (Fig.~\ref{LF_scg0018}). 
The same method described in Section~2 was adopted. A more detailed
analysis will be presented in a forthcoming paper.

The LF of NGC 92 exhibit a considerable flattening at the high luminosity end.
This effect is not shared by the companion galaxies, so it might be a real feature, 
rather than an artificial effect due to source blending.
When the highest luminosity H\,{\sc ii} regions are excluded from the fit of the LF,
we obtain a slope $\alpha$  = $-$1.73$\pm$0.11 consistent with the morphological 
type and in agreement, within errors, with the observed size distribution 
($\beta$ = $-$3.71$\pm$0.54).

Some excess of relatively high luminosity  H\,{\sc ii} regions is observed
also in ESO 194-G13, but we cannot exclude that this smaller effect is due
to source blending, especially in the central part of the galaxy, where the
spatial density of H\,{\sc ii} regions appears higher.
The observed size distribution is steeper than expected 
($\beta$ = $-$5.46$\pm$0.40), if the slope of the LF, $\alpha$  = $-$1.90$\pm$0.37,
is considered.

The LF of NGC 87 is based on the only 56 H\,{\sc ii} regions detected
with S/N $>$ 5$\sigma$. Also in this case the LF slope ($\alpha$ = $-$1.55$\pm$0.11), 
although consistent with the late morphological type of the galaxy (IBm pec), does not agree 
with the slope of the size distribution ($\beta$ = $-$3.57$\pm$0.73).

For the two remaining galaxies no H\,{\sc ii} region LF could be built. 
None the less, the H$\alpha$ images show interesting features.
NGC 88, apart from the bulk of emission stemming from its central part, shows
H$\alpha$ emission in two plume-like features.
NGC 89, besides hosting a Seyfert 2 nucleus, shows a ring of circumnuclear 
H\,{\sc ii} regions and two-sided extraplanar H$\alpha$ emitting features,
including a jet-like structure with a projected length of $\sim$ 4 kpc
whose nature is yet to be established (Fig.~\ref{LF_scg0018}).

\section{CG J1720-67.8}
\begin{figure}[hbt]
\centerline{
\hbox{
\includegraphics[width=6cm]{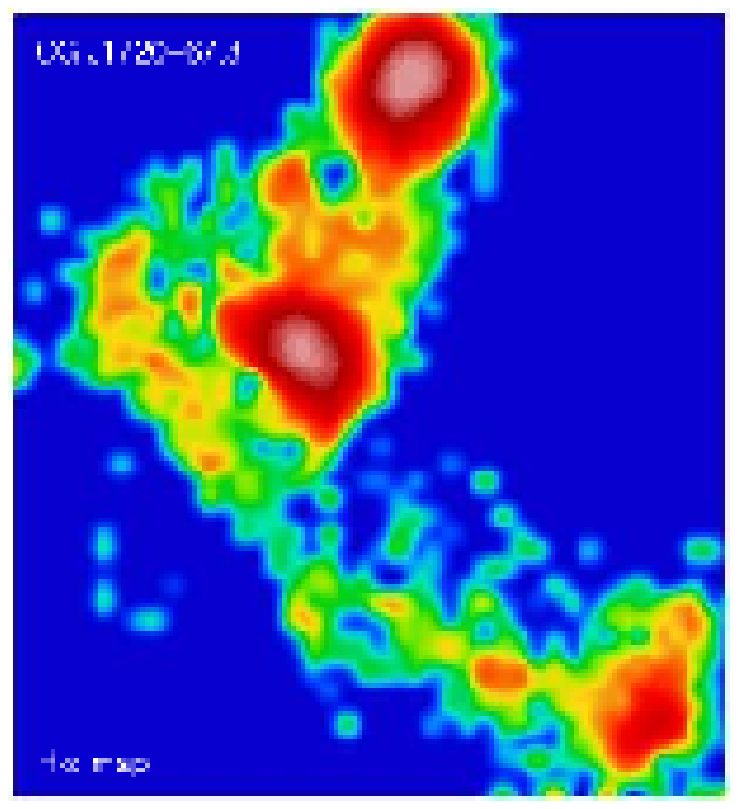}
\includegraphics[width=6cm]{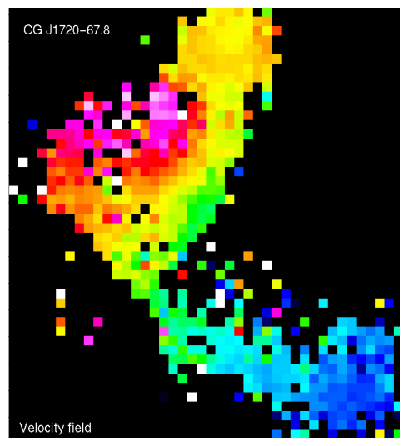}}}
\sidebyside
{\caption{Reconstructed H$\alpha$ map of CG J1720-67.8, based
on a mosaic of pointings with the integral field unit {\sc spiral}
mounted on the AAT.} \label{Hamap}}
{\caption{H$\alpha$ velocity field of CG J1720-67.8, reconstructed
from integral field spectra. Structures with lower (higher) radial 
velocities appear blue (red).}\label{vfield}}
\end{figure}

This extremely dense, low velocity dispersion compact group is
composed by two actively star-forming spirals and one lenticular galaxy with a 
small amount of central star formation \cite{T03b}.
A $\sim$ 29 kpc-long tidal tail hosts a number of candidate proto-tidal dwarf
galaxies \cite{T03c}. As shown in the H$\alpha$ map (Fig.~\ref{Hamap}) reconstructed
from integral field spectra obtained at the AAT 3.9 m telescope (see 
Temporin, Staveley-Smith, \&\ Kerber 2004 for a detailed description
and analysis of the data set), the star formation activity is spread all across
the group, with as much as 31 per cent of the entire H$\alpha$ emission stemming
from the long tidal tail. Such an activity is clearly related to the undergoing
merging process. The velocity field of the ionized gas (Fig.~\ref{vfield}) 
helps constraining the group dynamics and favours the idea of a prograde-retrograde
$\approx$ 200 Myr-old encounter between the two spiral galaxies.
While the estimated age of the tidal tail well agrees with the burst age of the two 
spiral galaxies, the knots in the tidal tail show evidence of more recent star formation
events (< 10 Myr), supporting the idea that the condensations of gas and stars
have formed under the action of self-gravity within the tail.

\section{Main results}
All three systems, 
although in different phases of interaction, present a clear
enhancement of star formation, which appears related to the
interaction process. This suggests encounter geometries 
favouring  the onset of bursts of star formation \cite{Mi93}.
The slopes of the H\,{\sc ii} region LFs mostly agree with
the expectations for the relevant morphological types
(Kennicutt et al. 1989), although some peculiarities are present.
Our main results on the individual systems are listed below.

%\subsection{Individual systems}
\emph{AM 1238-362.} The distribution of H\,{\sc ii} regions 
across the disks of the two face-on members seems related to the
interaction. Their H\,{\sc ii} region LFs, presented here for the
first time, considerably differ from one another, although the size 
distributions do not. The shift of the LF of Tol 1238-36.4 towards
high luminosities is possibly a consequence of the interaction process.
In this system we do not observe the tendency suggested by
Frattare, Keel, \&\ Laurikainen (1993) for interacting galaxies to
have flatter H\,{\sc ii} region LFs than normal galaxies.
Candidate intergalactic H\,{\sc ii} regions have been identified in
the space between the galaxies, however, spectroscopic confirmation
is needed. No obvious correlation is found between the
position of these sources and the distribution of H\,{\sc i} \cite{BW01}.

\emph{SCG 0018-4854.} All five galaxies show active star formation.
A preliminary analysis of the H\,{\sc ii} region LFs of three 
members of this compact group do show some peculiarities, 
especially a considerable flattening of the LF of NGC 92 at the
high luminosity end. A certain amount of flattening could be caused
by blending of H\,{\sc ii} regions, an effect that might become important
at a distance of 45 Mpc. However, since such an effect
was not observed in the other galaxies of the group, nor in 
AM 1238-362 (located at similar distance), we exclude this as the
only reason for the observed flattening. We plan to investigate separately the
LFs of the regions located in the tidal tail and in the galaxy body, 
where they are organized in circumnuclear rings.
Two of the galaxies appear to have H\,{\sc ii} region size distributions
steeper than expected, given the observed LFs. Possible reasons for
this effect are yet to be investigated.

\emph{CG J1720-67.8.} In this dense system, the merging process
is undoubtedly responsible for the ubiquitous star formation and the formation
of (candidate) proto-tidal dwarf galaxies. The velocity field of the ionized gas,
besides helping to constrain the interaction history of the group, shows an 
interesting feature, i.e. a $\sim$ 5 kpc offset of kinematic center with 
respect to the photometric center of the central spiral galaxy, in north-west
direction along its kinematic minor axis (Temporin et al. 2004). 
We speculate that this might hint at the presence of a common dark matter halo 
in the system.

\begin{acknowledgments}
In this work we made use of the {\sc HIIphot} algorithm available at the ftp site:
ftp://ftp.aoc.nrao.edu/staff/dthilker/HIIphot\_code/.
ST acknowledges financial support by the European Southern Observatory and is grateful for 
hospitality at the ESO-La Silla headquarter and at the Astronomy Department of Padova
University during the preparation of parts of this work. 
ST acknowledges financial support by the Austrian Science Fund (FWF)
under project no. P15065.
This work is based on data collected at the ESO-MPIA 2.2 m telescope equipped with EFOSC2, 
at the ESO-VLT-UT4 8 m telescope equipped with FORS2, and at the AAO-AAT 3.9 m telescope 
equipped with SPIRAL.
\end{acknowledgments}

\end{document}